\documentclass{eptcs}

\usepackage{amstext, amssymb, amsfonts, amsmath,mathbbol,eufrak}
\usepackage[mathscr]{euscript}

\newtheorem{example}{Example}

\begin{document}

\author{Gregor G\"ossler\institute{Univ. Grenoble Alpes, {\sc Inria}, CNRS, Grenoble INP, LIG, F-38000 Grenoble, France} \and Oleg Sokolsky\institute{University of Pennsylvania, Philadelphia, PA 19104, USA} \and Jean-Bernard Stefani\institute{Univ. Grenoble Alpes, {\sc Inria}, CNRS, Grenoble INP, LIG, F-38000 Grenoble, France}}

\providecommand{\event}{CREST 2017}
\def\titlerunning{Counterfactual Causality from First Principles?}
\def\authorrunning{G. G\"ossler, O. Sokolsky, and J.-B. Stefani}

\title{Counterfactual Causality from First Principles?\thanks{This work was supported by the {\sc Causalysis} associate team funded by {\sc Inria}.}}

\maketitle

\begin{abstract}
  In this position paper we discuss three main shortcomings of
  existing approaches to counterfactual causality from the computer
  science perspective, and sketch lines of work to try and overcome
  these issues: (1) causality definitions should be driven by a set of
  precisely specified requirements rather than specific examples; (2)
  causality frameworks should support system dynamics; (3) causality
  analysis should have a well-understood behavior in presence of
  abstraction.
\end{abstract}

\section{Introduction}

Counterfactual reasoning has multiple applications to forensic
analysis of failures in safety-critical systems. Modern embedded and
cyber-physical systems are characterized by a large number of
concurrent components with multiple interactions between them.
Furthermore, physical environment of such systems adds to the
complexity of dynamics of system executions, and some of the physical
interactions may not be directly observable. Consider an example from
the medical domain~\cite{pajicTII2012}. A patient is being treated for
pain using a medication delivered by an infusion pump. To prevent
overdoses, which can be fatal, the system is equipped with a safety
interlock that stops the pump if a dangerous condition is detected
through vital sign sensors, such as blood oxygenation and pulse rate.
If an overdose occurs, causality analysis faces several challenges.
The pump may be infusing medication at a slightly higher rate than
programmed, but was it the cause?  Infusion is a continuous process,
and duration of the infusion is just as important factor as the
infusion rate. Proper abstraction of such an interaction is important
for the analysis. Similarly, the interlock may not have detected the
overdose symptoms in time, but was it the problem of the algorithm
used by the interlock or the fact that the patient had unusually high
sensitivity to the drug. Complex dynamics of the physiological effects
of pain medication need to be taken into account, since they directly
affect how quickly the patient gets overdosed.

In the rest of this paper, we present three research directions in
counterfactual causality analysis that may help, in the future, to
address challenges posed by this motivating example.  First, we argue
that causality definitions should be driven not by individual examples
but by a set of precisely specified requirements and discuss what
these requirements may include.  Second, we argue that support for
system dynamics and temporal relationships should be included in the
causality framework.  Note that, by themselves, the techniques
discussed may not be sufficient to incorporate continuous dynamics
that the example needs.  However, the third research direction deals
with the use of abstractions in causality analysis. Such abstraction
techniques as discretization of continuous signals may allow us to
eventually handle the full complexity of causality analysis in
cyber-physical systems.

\section{Escaping the TEGAR}

Research on counterfactual causality analysis has been marked, since
its early days \cite{Hume1739}, by a succession of definitions of
causality that are informally (in)validated against human intuition on
mostly simple examples. Let us call this approach TEGAR, {\em textbook
  example guided analysis refinement}. TEGAR contrasts with
mathematics and natural sciences building knowledge on theorems and
proofs, in that most work on causation lacks formal properties against
which the definitions are tested. As a result, TEGAR suffers, as
pointed out in \cite{glymour2010actual}, from its dependence on the
tiny number and incompleteness of examples in the literature and the
lack of stability of the intuitive judgments against which the
definitions are validated. This absence of formal tools for evaluating
theories of causation is not primarily owed to a lack of
formalization: at least since the works of
\cite{spirtes2000causation,Pearl2000}, different formal definitions of
causality have been proposed. Among the most influential definitions
of counterfactual causality are Lewis' possible world semantics
\cite{Lewis1973b,Lewis1986,Lewis00} and Pearl's and Halpern's actual
causality \cite{Pearl2000,HalpernPearl2005,DBLP:conf/ijcai/Halpern15}.
Both definitions have undergone a series of refinements in order to
match human intuition on additional examples that were proposed to
challenge them. One may doubt that this is the end of the story. It is
interesting to note that the understanding of causality and
explanations in natural sciences faces a similar lack of objective
referential, as witnessed by the dispute reedited in
\cite{CausalityInPhysics1990}.

We believe that a more constructive, reproducible approach to design
definitions of counterfactual causality is needed. A first step
towards this goal would be to formally define a family of requirements
--- that are as agnostic as possible of concrete models of computation
--- on counterfactual causality, rather than a mere set of competing
concrete definitions.

Some efforts to axiomatize counterfactual reasoning have been made. On
structural equations models \cite{Pearl2000} (SEM),
\cite{springerlink:10.1023/A:1009602825894} introduces three
properties that hold in all (recursive and nonrecursive) SEM, two of
which characterize manipulation (Pearl's {\em do} operator) in the
recursive case. \cite{journals/jair/Halpern00} generalizes these
results to an axiomatic characterization of the classes of
non-recursive SEM with unique solutions, and arbitrary SEM. With the
goal of using counterfactual causality for fault ascription --- that
is, blaming a system failure on one or more component faults ---,
\cite{GoesslerStefani2015} proposes general constraints on
counterfactuals that are sufficient to entail correctness and
completeness. Similarly, a general definition of actual causality is
proposed and then instantiated in
\cite{journals/ijar/BeckersV16}. While none of these axiom systems is
strong enough to characterize more than basic properties of
counterfactual reasoning, we believe that there is still room for
progress.

Let us have a closer look why current accounts of counterfactual causality are
not satisfactory. The impact of modeling choices on counterfactual analysis has
long been recognized, see for instance \cite{DBLP:journals/corr/abs-1106-2652}. In our
eyes, one of the most critical shortcomings of state-of-the-art SEM-based
approaches is the dependence of the result on the structure of the model.

\begin{example}[Lamp \cite{weslake2015partial}]
  Consider three variables $A$, $B$, $C$ ranging over $\{-1,0,1\}$.
  Lamp $L_1$ is on whenever any two of the variables share the same
  value. Lamp $L_2$ is on whether there is a value among $\{-1,0,1\}$
  that is different from the values of all three variables. The
  structural equations are as follows:
  \begin{align*}
    L_1 \; = \; & \; (A=B \; \vee \; B=C \; \vee \; A=C) \\[1ex]
    L_2 \; = \; & \; (N_{-1} \; \vee \; N_0 \; \vee \; N_1) \quad\mbox{where} \\
    N_i \; = \; & \; (A\neq i \, \wedge \, B\neq i \, \wedge \, C\neq i), \quad i\in\{-1,0,1\}
  \end{align*}
  In the actual world, the state is $A=1$, $B=C=-1$, $N_{-1}=0$,
  $N_0=1$, $N_1=0$, and $L_1=L_2=1$. Halpern's modified definition of
  actual causality \cite{DBLP:conf/ijcai/Halpern15} considers each of
  $B=-1$ and $C=-1$ as a cause of $L_1=1$ and of $L_2=1$. However, it
  also considers $A=1$ as a cause of $L_2=1$.\footnote{The previous
    definitions of actual causality \cite{Pearl2000,HalpernPearl2005}
    consider each of $A=1$, $B=-1$, $C=-1$ as a cause of both $L_1=1$
    and $L_2=1$.}  Intuitively, this is due to the fact that there is
  a contingency --- namely, holding $N_1$ at zero --- under which
  switching $A$ from $1$ to $0$ switches off $L_2$. Thus, the
  resulting causes for $L_1=1$ and $L_2=1$ differ even though the
  definitions of $L_1$ and $L_2$ are logically equivalent.
\end{example}

This example brings us to a first set of points we want to make.

First, the result of causality analysis should depend on the semantics
of the model but not its syntax. An intuitive motivation for this
requirement is that the analysis should be ``objective enough'' so as
to determine causality independently of how the story is told. More
importantly, the formal motivation is that searching for a theory of
causation that allows us to reason about equivalence and refinement of
models is hopeless as long as semantically equivalent models are
distinguishable.

Second, we need formalization of counterfactual causality based on
{\em first principles}, similar to the approach of
\cite{DBLP:journals/ai/Reiter87}, in the sense that the formalization
is constructed from general requirements. As in the design processes
in most engineering disciplines, the development of a definition of
causation should be performed top-down, starting from the question of
{\em what formal requirements the definition should meet},
independently of concrete modeling frameworks. An example of such
requirements is robustness of causation under equivalence of models,
for a given definition of equivalence. Directly focusing on the
question ``how to implement it?'' is likely to narrow down the design
space prematurely and require ``debugging'' of the definitions, as
discussed above. In turn, such a ``specification'' of counterfactual
causality should help us in answering questions such as:
\begin{itemize}

\item How to design a counterfactual analysis satisfying the
  requirements?

\item Can we obtain the same analysis result with other tools than
  counterfactual analysis? What are properties of interest that are
  satisfied only by counterfactual analysis?

\item Is counterfactual causality analysis inherently NP-complete (as
  Halpern and Pearl's actual
  causality) \cite{DBLP:journals/ai/EiterL02,DBLP:conf/ijcai/Halpern15}?
  
\end{itemize}

\section{Native Support for System Dynamics}

It has been pointed out in \cite{glymour2010actual} that Halpern and
Pearl's definitions of actual causality, based on SEM over
propositions, poorly support reasoning about state changes.  Other
limitations of SEM --- in particular, their inability to distinguish
between states and events, and between presence and absence of an
event --- have also been noted e.g.\ by Hopkins and Pearl
\cite{HopkinsP07}, and several other formalisms have been suggested
for supporting reasoning about causal ascription (see for instance
\cite{BenferhatBCNDSKNNPS08}). Counterfactual definitions of necessary
causality for behaviors over time have been proposed for biochemical
reactions in \cite{danos_et_al:LIPIcs:2012:3866} and similarly for
programs in \cite{conf/csfw/Datta0KSS15}, and for fault ascription in
component-based systems in \cite{GoesslerLeMetayer2015}; some works
define variants of actual causality on models of execution traces
\cite{DBLP:journals/fmsd/BeerBCOT12,DBLP:conf/safecomp/KuntzLL11}.

Apart from the modeling infelicities of SEM, a key point is that
models that allow finitary descriptions of systems dynamics are
essential for conducting actual cause analysis. In particular since
counterfactual executions may be unbounded it may be necessary to
explore a prefix of the counterfactuals whose length is not bounded a
priori, in order to evaluate the property. For instance, 
a system dynamics can be represented by a set of traces or some
sort of automata, and actual cause analysis for a property violation during some execution
can consist in constructing 
sets of traces or automata executions that avoid a particular set of violating states
but keep at least the antecedent part of the original execution.
In order to effectively
construct and analyze these counterfactual executions, we then need a symbolic representation,
along with symbolic formulations of the counterfactual construction
and analysis. Symbolic approaches to causality checking have been
proposed e.g.\ in \cite{conf/spin/BeerHKLL15} for Halpern and Pearl's
actual causality and in \cite{RV2013,emsoft14} for fault ascription in
real-time systems; except for \cite{emsoft14} they rely on generating
and analyzing bounded counterfactuals.

For systems dynamics, the notion of coalgebra \cite{Jacobs2016,Rutten00}
provides a systematic setting, generalizing notions of transition systems
that include e.g.\ many variants of probabilistic and stochastic transition
systems \cite{Sokolova11}, as well as hybrid transition systems \cite{Neves2016}.
Following the (hyper)set-based formulation of \cite{Barwise96Circles}, a system
can be described coalgebraically as a possibly infinite, mutually recursive, set of equations of the form
$x = F(x)$, where $F$ is some operator on sets, and $x$ some variable. 
For instance, the standard notion of
(finitely branching) labelled transition system is given by operator $F$ defined
as $F(X) = \mathscr{P}_f(A \times X)$ where $\mathscr{P}_f (S)$ denotes the set of finite
subsets of some set $S$, $X$ is the set of (state) variables and $A$ is the set of labels.
One benefit of the coalgebraic approach is its
generality. For instance, many different variants of transition systems,
including timed, quantitative, and stochastic ones, are instances of coalgebras.
Our contention is that it could be beneficial to develop causality analysis 
in an abstract coalgebraic framework, if only to identify abstractions and 
constructions (e.g.\ for counterfactuals) that apply generally irrespective of the
actual details of the chosen operators. \cite{GoesslerStefani2015} provides an example
of counterfactual analysis developed in an abstract setting -- that
of configuration structures, which can be understood as a general model for
concurrent system executions or unfoldings. It seems to us that general notions of causality
and counterfactuals should not depend on the specifics of system or transition system models.
Rather, we expect that at least general constraints on counterfactual construction and causal dependencies
can be obtained for abstract system models and properties. 
For instance, a general notion of behavior and bisimulation can be defined for coalgebraic systems \cite{Jacobs2016,Rutten00},
that does not depend on the specifics of the chosen operator. Obtaining similarly
abstract characterizations of causal dependencies or counterfactuals would be of enormous benefit.

Once we allow for unbounded executions also in the actual world ---
that is, the observed execution, --- {\em incremental} causality analysis
becomes an issue.  Many causality analysis techniques operate on the observed prefix of the execution at the point when an event of interest, such as a failure, is discovered.  If the analysis relies on explicit construction of counterfactuals, there is a danger of repeating the same work for different counterfactuals.  Moreover, if the analysis has to be performed multiple times over an evolving execution, redundant efforts are even more likely.  In this case, incremental analysis can keep partially constructed counterfactuals, hopefully in a symbolic form, and update them as the next observation from the execution arrives.  While this approach may not reduce complexity of causality analysis, it may amortize the cost over a long-running execution and reduce analysis latency, once an event of interest is observed.

With this vision, the partial, evolving counterfactual would allow us to answer
the question, ``if the event of interest is to happen in the next step, what
would the causes be?''  When there are multiple events of interest --- for instance, multiple ways for a failure to occur, --- the danger is that the incremental analysis would incur additional cost with bookkeeping for events that never occur.

\section{Causation and Abstraction}

Important applications of causality analysis include the construction
of concise explanations for observed behaviors
\cite{DBLP:journals/fmsd/BeerBCOT12,danos_et_al:LIPIcs:2012:3866}, and
establishing liability \cite{DBLP:journals/cacm/MetayerMMPFTCH11}.
Preconditions for causality analyses to be applicable and sound are
(a) availability of the necessary observations to determine causality,
and (b) consistency between the model on which the analysis is
performed, and the implementation generating the
observations. Requirement (a) is addressed by ensuring {\em
  accountability} \cite{DBLP:conf/ccs/KustersTV10} with respect to
causality analysis, that is, constructing systems in such a way that
all information necessary to elucidate the causes of events of
interest is logged. We believe that accountability with respect to
causality analysis should become a design requirement for new designs
of safety-critical systems.

Surprisingly, requirement (b) has received little attention in the
computer science community so far. However, using counterfactual
analysis on hand-crafted models of causal dependencies to determine
the causes of a system failure is much like modeling a critical system
in one formalism and then implementing it in another one from scratch
--- the semantic gap between the model and the actual system makes it
difficult to ensure that the former is faithful with respect to the
latter. Software design has been formalized as a series of refinements
from a high-level specification down to the implementation; for the
design of cyber-physical systems, numerous discrete abstractions of
the continuous dynamics have been proposed, see
e.g.~\cite{tabuada2009}. Theories of causation should therefore be
able to track causation through these levels of abstraction and
refinement, for instance, to verify causation on a small abstract
model and then refine potential causes identified on that level. To
this end, theories of causation should have a well-defined behavior
under abstraction and refinement, such as correctness (any cause in
the abstract model is refined into a cause in the refinement) or
completeness (the abstraction of any cause in the refinement is also a
cause in the abstract model) of abstraction. One can go even further
and ask how causality meshes with system equivalences. The standard
benchmark for system equivalences is contextual equivalence: given
some notion of observable and some notion of system execution, two
systems are equivalent when, placed in the same context, they have the
same observables and the same executions.  It seems to us plausible to
ask of a notion of causality to be robust with respect to contextual
equivalence: if causal analysis in a complex system $S[A]$, where
$S[\cdot]$ is a context for subsystem $A$, yields a certain result,
then the same analysis performed on $S[B]$, where $B$ is contextually
equivalent to $A$, should yield the same result (e.g.\ pinpointing
some observable event in $A$ or $B$ as the actual cause of some
property violation).  We are not aware of any work studying
abstraction, refinement, or robustness, in the SEM framework.

Finally, deriving an implementation by refining an abstract
specification usually implies that the abstract model encompasses some
non-determinism. In order to support multiple levels of refinement,
theories of causation have to be able to cope with this
non-determinism.

\bibliography{pp}

\end{document}